\documentclass{article}

\usepackage{arxiv}

\usepackage[utf8]{inputenc} % allow utf-8 input
\usepackage[T1]{fontenc}    % use 8-bit T1 fonts
\usepackage{hyperref}       % hyperlinks
\usepackage{url}            % simple URL typesetting
\usepackage{booktabs}       % professional-quality tables
\usepackage{amsfonts}       % blackboard math symbols
\usepackage{nicefrac}       % compact symbols for 1/2, etc.
\usepackage{microtype}      % microtypography
\usepackage{lipsum}
\usepackage{graphicx}
\usepackage{multirow}
\usepackage{xcolor}
\usepackage{listings}
\graphicspath{ {./images/} }

\title{The CARFAC v2 Cochlear Model in Matlab, NumPy, and JAX}

\author{
 Richard F. Lyon \\
  Google Research\\
  \texttt{dicklyon@ieee.org} \\
   \And
 Rob Schonberger \\
  Google Platforms and Services\\
  \texttt{robsc@google.com} \\
  \And
 Malcolm Slaney \\
  Google Research\\
  \texttt{malcolm@ieee.org} \\
  \AND
 Mihajlo Velimirović \\
  Google Deepmind\\
  \texttt{mvelimirovic@google.com} \\
   \And
 Honglin Yu \\
  Google Research\\
  \texttt{honglinyu@google.com} \\
}

\begin{document}
\maketitle
\begin{abstract}
The open-source CARFAC (Cascade of Asymmetric Resonators with Fast-Acting Compression) cochlear model is upgraded to version 2, with improvements to the Matlab implementation, and with new Python/NumPy and JAX implementations---but C++ version changes are still pending.  One change addresses the DC (direct current, or zero frequency) quadratic distortion anomaly previously reported; another reduces the neural synchrony at high frequencies; the others have little or no noticeable effect in the default configuration.  A new feature allows modeling a reduction of cochlear amplifier function, as a step toward a differentiable parameterized model of hearing impairment.  In addition, the integration into the Auditory Model Toolbox (AMT) has been extensively improved, as the prior integration had bugs that made it unsuitable for including CARFAC in multi-model comparisons.
\end{abstract}


\section{Introduction}
With the CARFAC cochlear model \cite{lyon2011cascades,lyon2017human} being used in several research projects \cite{xu2018fpga,xu2019binaural,xu2021biologically,islam2022noise}, and being included in some multi-model comparisons \cite{saremi2016comparative} and omitted from others \cite{osses2022comparative}, we have in recent years found a few things worth improving.  This report documents recent fixes and changes released on GitHub \cite{carfac2012github}.

\section{Extended Release Notes}
\label{sec:notes}

This section describes changes made to create version 2 of CARFAC, including parallel changes in the MATLAB, NumPy, and JAX versions, but not yet in the C++ version.  We will refer to this update on GitHub as CARFAC v2 (or as lyon2024, as opposed to lyon2011, in the Auditory Model Toolbox adaptation \cite{majdak_amt_2022}).  The v2 changes includes a RELEASE\_NOTES.md file with 10 numbered changes that we expand on here in correspondingly numbered subsections.

References to ``the book'' herein are to the Lyon 2017 book \cite{lyon2017human} that describes the CARFAC model in detail.

\subsection{AC Coupler Moved: Highpass Filter Now at BM Output}
\label{sec:coupler}

To suppress DC and very low frequency response in the BM output, the 20 Hz AC-coupling (highpass) filter that was the first part of the inner hair cell (IHC) has been moved into the cascade of asymmetric resonators (CAR).  The basilar membrane (BM) output of the CAR is now the output of that filter.  This filter roughly models helicotrema reflection that ``shorts out'' zero-frequency (DC) BM response, as suggested previously \cite{saremi2018quadratic}.

Motivated by observations of anomalous DC energy in the BM output  \cite{saremi2018quadratic}, this change suppresses DC while keeping the default OHC parameters that lead to plausibly realistic amounts of low-frequency quadratic distortion at audible frequencies, which may be important in musical pitch perception. \cite{patterson2013modelling,pressnitzer2001distortion}.

\begin{figure}[!ht]
  \centering \includegraphics[width=3.5in]{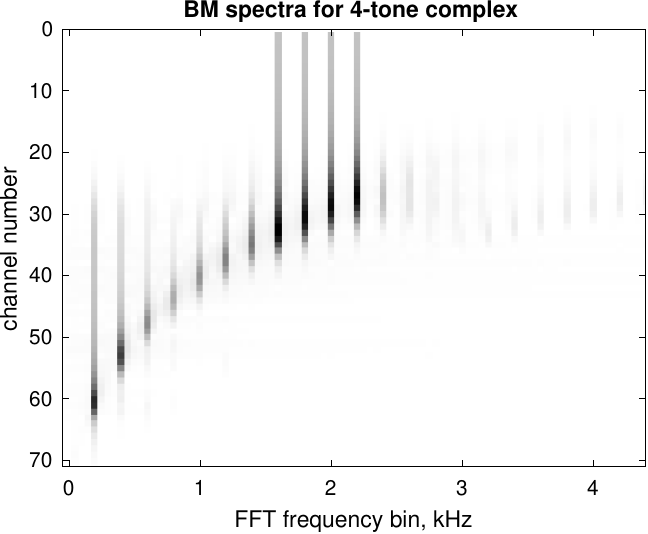}
    \caption{{\bf Spectral view of propagating distortion products.} This FFT analysis of the BM outputs, with four-tone waveform input, shows no propagating DC (0 Hz) component, unlike the book's Figure 17.7.  Other low-frequency quadratic distortion tones (e.g. 200, 400, and 600 Hz) are still prominent, as are cubic distortion tones (e.g. 1000, 1200, 1400, and 2400 Hz}.
  \label{fig:bm_fft_v2}
\end{figure}

The IHC output (the neural activity pattern, NAP) is not affected by this move of the filter across the CAR--IHC boundary, but the BM output is: recomputing the book's Figure 17.7 shows that the stripe at 0 frequency disappears, as shown in Figure \ref{fig:bm_fft_v2}.

With the AC-coupling filter relocated, there is reduced very-low-frequency distortion effect at the BM output.  Of course, this highpass filter reduces the BM response not just to low-frequency quadratic distortion, but also to low-frequency input sounds, with a half-power corner at 20 Hz, so some frequency-response test data had to be updated.  Roughly, this filter models the shunting of low-frequency pressure differences through the helicotrema, whereas when it was in the IHC it might have been seen as modeling viscous coupling of fluid to cilia.  Perhaps two such AC-coupling filters, one in each place, might make sense, but that's not currently supported.  It might be more realistic to raise the corner, perhaps to 30 or 40 Hz; the parameter for that is now in the CAR params, not the IHC params.

\subsection{Open-Loop Bug Fix}
\label{sec:olbug}

In v1, if the mode was changed to open-loop while the AGC (automatic gain control) feedback-controlled filter parameter were changing, the ramping that was intended to interpolate to the new value would continue, extrapolating beyond the intended value.  With this bug fix, the ramping is stopped when calling \texttt{CARFAC\_Run\_Segment} in open-loop mode.

The open-loop bug was fixed in the C++ version in 2017, in support of tests there, but wasn't fixed in the Matlab until 2022.  There was previously \texttt{CARFAC\_Run\_Open\_Loop} as a workable alternative in Matlab.

\subsection{Linear CAR Option}
\label{sec:linrun}

The linear mode simply sets the OHC (outer hair cell) nonlinear function output to always 1.0, to make each CAR stage linear if the AGC feedback is not changing (that is, if in open-loop mode).  This is useful to measuring things like impulse responses to characterize the system with linear-systems concepts, e.g. in tests.

As an alternative, there was also \texttt{CARFAC\_Run\_Linear} to skip all the OHC, AGC, and IHC processing.  The improved parameterized \texttt{CARFAC\_Run\_Segment} is a better way to conditionally run in linear mode.  The old functions have not been removed.

If the CAR is not run open loop, there will still be gain adaptation via the relative undamping parameters in the stages, yielding compression and some odd-order distortion.  Either open or closed, the relative undamping will include the effect of outer hair impairment if that new feature is used.

\subsection{Two-Capacitor IHC Model}
\label{sec:ihc}

We removed the previous (not used by default) two-capacitor IHC model, and replaced it by a new one that allows interpreting one of the capacitor voltages as a receptor potential.  This new two-cap version is the default in v2; it results in rather minor differences in the IHC output (including a somewhat reduced peak-to-average ratio in the NAP output), and via AGC feedback also very tiny differences in the BM output.

This is a key functional difference in v2, so we characterize its effect a bit more here.

\begin{figure}[!ht]
  \centering \includegraphics[width=5in]{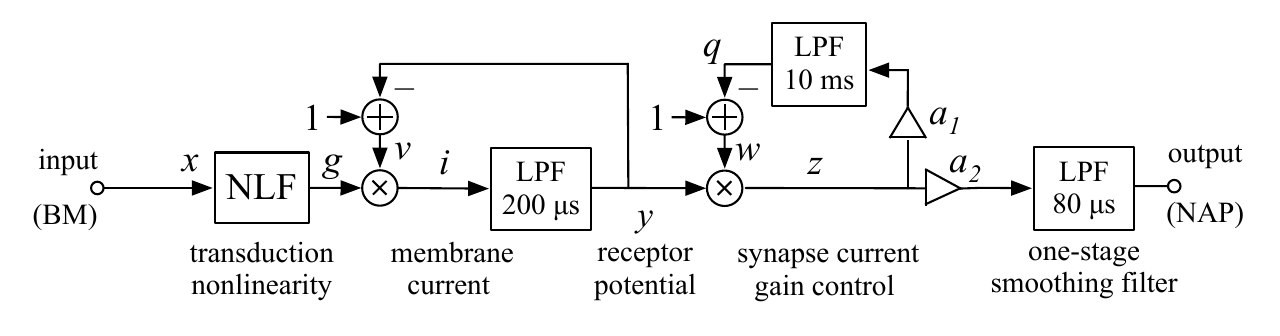}
    \caption{{\bf The two-cap IHC model.} The dynamic state of the IHC is in the three LPF blocks, which are unity-gain one-pole smoothing filters.  The first one smooths the receptor potential, and is the main effect contributing to reduced synchrony at high frequencies; the last one further reduces the synchrony.  The middle LPF represents the reservoir of neurotransmitter that can deplete and recover on a time scale of milliseconds; it does not cause any smoothing in the forward pass, like the one in the one-cap model and unlike the first LPF here.}  
  \label{fig:ihc_v2_block}
\end{figure}

\begin{figure}[!ht]
  \centering \includegraphics[width=3.5in]{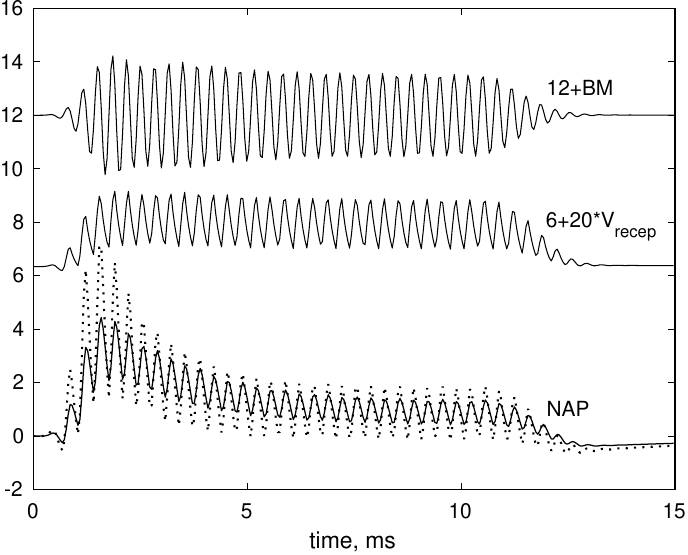}
    \caption{{\bf IHC model tone-burst responses.} Responses of the IHC to 3 kHz 10 ms tone burst (without ramping at onset or offset, starting at time 0, at $-40$ dBFS, channel 23 in default CARFAC, with CF near 3 kHz).  Top: BM response, the IHC input (this is from v2, but v1 is not noticeably different), offset to +12.  Middle: the receptor potential of the two-cap IHC model, multiplied by 20 and offset to +6.  Bottom: NAP response, the IHC output, with the corresponding v1 output dotted.  Note that the receptor potential shows a subtle onset sharpening, but not much overshoot.  The v2 NAP still shows a strong onset emphasis, but about half the AC component, or vector strength, of the v1 model at 3 kHz; the difference is much less at lower frequencies.}  
  \label{fig:ihc_compare}
\end{figure}

Here, ``capacitor'' refers to a state variable whose change affects the signal multiplicatively, e.g. as the voltage driving the receptor current, or the neurotransmitter store driving the synaptic release at the output.  Other modelers have used terms such as ``reservoirs'', ``stores'', or ``pools'' for such state variables, as in ``modeled by introducing an additional transmitter reservoir between the
global store (factory) and the free-transmitter pool.''\cite{meddis1988simulation}  The first two lowpass filters (LPFs) in Figure \ref{fig:ihc_v2_block} correspond to the two capacitor states.  The figure is somewhat abstracted from both the circuit and the code, but should accurately represent the two-cap IHC model as implemented, except that it doesn't show that we subtract the steady-state silence output value at the end, to get a quiescent output of zero as in v1 (which was possibly not a good idea, but we didn't change it for compatibility reasons).

\begin{figure}[!ht]
  \centering \includegraphics[width=3.5in]{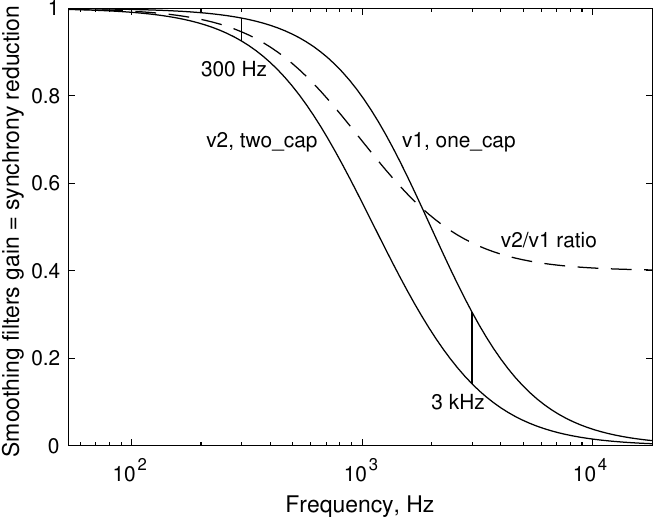}
    \caption{{\bf IHC smoothing comparison.}  The longer smoothing time constant in v2's two-cap IHC model yields a (linearized, small-signal) magnitude transfer function that is lower than the one for v1's one-cap model.  Their ratio is plotted (dashed), and lines are drawn connecting the points corresponding to the 3 kHz tone plotted in Figure \ref{fig:ihc_compare}, as well as the 300 Hz tone that is shown in the book, where the synchrony reduction is minor. This more severe loss of synchrony in v2 may be more realistic.}
  \label{fig:ihc_smoothing}
\end{figure}

The v2 two-cap IHC's neural activity pattern (NAP) output has less synchrony to high frequencies, as shown in Figure \ref{fig:ihc_compare}, with smoothing time constants of $200 \mu$s and $80 \mu$s, compared to the two 80 $\mu$s smoothers of the one-cap model that was the default in v1.  The slower first time constant, which determines the amount of synchrony at the receptor potential, may be more accurate for humans and guinea pigs, but too slow for cats. These can be changed in the design parameters, if calibrating closer to particular experimental data is desired.  The v1 and v2 default small-signal transfer functions are compared, and their ratio plotted, in Figure \ref{fig:ihc_smoothing}.

\subsection{IHC Model Selection}
\label{sec:ihc_choice}

The function \texttt{CARFAC\_Design} now takes an optional last arg, keyword \texttt{one\_cap} or \texttt{just\_hwr} to change from the default \texttt{two\_cap} inner hair cell model.  This makes it easier to get back to the old one-cap IHC model if desired, and easier to make tests that compare them.

Switching a CARFAC instance between IHC types dynamically is not supported, since they do not have compatible state variables.

\subsection{Delay-per-Stage Option}

The CAR has minimum-phase response from the input to each BM output channel.  To achieve this, each stage at each sample time needs to use the immediately computed output of the stage before, which means those outputs cannot be computed in parallel. A sequential ``ripple'' across channels is used in the CAR update step, after most of work has been done by channel-parallel array operations.  By backing off on being minimum phase, adding a one-sample delay between stages, we can allow stage inputs to depend only on outputs from previous time steps.  This change removes the need for the ripple, allowing the whole CARFAC to be run with channel-parallel array operations.

Therefore, we now optionally include an extra sample delay per cascaded resonator stage.  For compatibility, this new option is off by default. To enable it in Matlab, set the \texttt{use\_delay\_buffer} field in the designed CARFAC struct, and the value will be copied into the \texttt{CAR\_params} in \texttt{CARFAC\_Run\_Segment}; if you're running the CAR through some other interface, consider whether that needs to be set.  In the NumPy version, it is an optional parameter to \texttt{design\_carfac}, and in JAX a slot in the \texttt{CarDesignParameters} class.

Tests verify that when the option is enabled, the outputs are staggered by one sample per stage as expected, and that the effect on short-time mean outputs is negligible, even though there is a small change in the timing of AGC feedback across channels (recall that the AGC feedback changes slowly, updating every 8 samples, so one-sample timing changes don't do much).

\subsection{Simplification of Stage Gain Updates}

The AGC feedback adaptation that moves the poles and zeros of the asymmetric resonator stages also changes their DC gain a little, so we update the gain factor between cascaded stages of the CAR to compensate, keeping the low-frequency tail of the transfer function stable.  It's a small effect, less than 1 dB per stage, but without compensating it, there will be an unrealistically large effect of suppression of very low frequencies by very high frequencies.  

In v1, we computed the exact DC gain needed, using a ratio of polynomials whose value varies over a large range across the channels, which was problematic for fixed-point FPGA implementations, where high-dynamic-range divides are expensive.  In the book, Figure 16.3, we had shown that the needed adjustment is nearly quadratic in damping factor, so we leveraged that in v2.  In particular, rather than using the small-damping approximation, we fitted a parabola to the values at the 0.0, 0.5, and 1.0 values of each channel's local undamping variable, so it remains accurate to about 0.01 dB over the entire range of adjustment, in all channels.

Tests verify that the result is not significantly different from the v1 result.  The old way is no longer supported.

\subsection{AGC Spatial Smoothing Simplification}

In v2, we removed the internal options to implement the AGC loop's spatial smoothing filter using a 5-point FIR filter or a forward/backward IIR filter, leaving just the 3-point FIR implementation, which matches the description and diagram in the book's Figure 19.6.  If, with the specified design parameters for time constants, sample rates, and decimation factors, the requested smoothing amounts are more than the 3-point FIR filter can achieve, the first AGC decimation factor will be reduced instead of switching to a different filter architecture.  Both the design code and the run-time code are thereby simplified.  This change simplifies ports to FPGA, JAX, etc.

There are still limits on the feasible design parameter space, if large amounts of spatial coupling are requested and the sample rate is not high enough to run the 3-point FIR filter enough times within a time constant to achieve the desired diffusion distance.  There is no change with the default design parameters (22050 Hz sampling rate, first AGC stage decimation by 8, and default AGC filter time constants, spreads, and basalward shifts described in the book).

\subsection{Outer Hair Cell Impairment}

We added a provision for modeling sensorineural hearing loss via the \texttt{ohc\_health} coefficient vector (a health value between 0 and 1 per channel, a final factor in the relative undamping, defaulting to 1).

To model a reduction in OHC activity associated with sensorineural hearing loss, we added a vector of \texttt{ohc\_health} coefficients---one number per stage, where a value of 0 means passive response and 1 means the default healthy level of activity.  These coefficients multiply the computed linearized undamping values in closing the AGC loop, before the stage gain computation.  The OHC's nonlinear function (NLF) multiplicatively reduces the undamping below the linearized value, whether impaired or not.  

The change in peak gain or sensitivity between the healthy and passive cases is frequency dependent, with a maximum around 50 dB.  Slightly negative values of \texttt{ohc\_health} can be used to model more impairment, increasing the damping above its nominal maximum, but then the NLF effect will be anomalous; it might be better to go back and change the CAR design to have a larger maximum damping value to correspond to the passive or dead cochlea; default max \texttt{zeta} is 0.3; increasing it to 0.35 or so would allow modeling a greater impairment.  Alternatively, recognizing that not all impairment is due to OHC dysfunction, we expect to add more mechanisms in the future to model synaptopathy, or loss of effective auditory nerve fibers, via an added stage or an elaboration of the IHC model.

Tests are included to show that setting \texttt{ohc\_health} to zero over a basal part of the cascade results in greatly reduced response to high frequencies and not much reduction of response to low frequencies.  The effect is not calibrated, but provides a vector of values that can be optimized to match an individual's audiogram or test results.

\subsection{Extensive Tests}

We have added test routines to the open-source code.  We have used these to help us keep the Matlab, NumPy, and JAX versions synchronized, and to facilitate updating the C++ version as well, in case someone wants to work on that.  The tests also produce plots, e.g. comparing the one-cap and two-cap IHC models at 300 Hz and 3000 Hz, showing significant but not huge differences.

\section{AMT Integration}
\label{sec:amt}

These changes are also propagated to the Auditory Model Toolbox, with names of files and functions changed to use the \texttt{lyon2011} and \texttt{lyon2024} prefixes.  The lyon2011 version includes some of the fixes, and corrected integration into AMT, and represents v1, while lyon2024 represents v2, the preferred version going forward.

One reason CARFAC v1 was not included in \cite{osses2022comparative} is that to the authors it was not clear how to get the auditory nerve or receptor potential output in a form that could be compared with other models. In CARFAC v2, we exposed more intermediate outputs, so that in v2 there is an analogue to the IHC receptor potential.

The code is currently in the lyon2024 branch of \href{https://sourceforge.net/projects/amtoolbox/}{\texttt the AMT repository}, and it should be propagated to the main branch once AMT version 1.6 is released.

\section{NumPy Version}
\label{sec:numpy}

A new Python/NumPy version of CARFAC was developed as a fairly literal transliteration of the Matlab, and tests were developed simultaneously as a way to make sure it was correct.  Variable names and function names were kept closer to the Matlab names than we did in the C++/Eigen version.  The NumPy version has been kept in sync with v2 changes that were being developed in the Matlab version, and the tests from the NumPy version were back-ported to the Matlab version.  The Matlab tests were extended to print the ``golden data'' used in some Python tests.

Typically, the Matlab and NumPy versions use 64-bit double-precision floating point (doubles), while JAX uses 32-bit single-precision (floats).  Some tolerances are adjusted in the tests as needed for comparisons to pass.  Typically the NumPy and Matlab versions match to better than a difference of $10^{-6}$, or better than 4 significant digits. The JAX versions match to a difference between $10^{-6}$ and $10^{-3}$, or better than 2 significant digits.

\section{JAX Version}
\label{sec:jax}

JAX \cite{jax2018github} is a relatively new, high-performance numerical computing framework for Python, which targets, but is not limited to, machine learning research. It can automatically compute the gradients of composed mathematical functions and just-in-time (JIT) compile the Python code into machine code optimized for accelerators (GPUs and TPUs). Having a JAX version of CARFAC allows us to fine-tune CARFAC's parameters through algorithms such as gradient descent, and makes it straightforward to let CARFAC and artificial neural network models work and be trained together, possibly on accelerators.

Although JAX reuses a NumPy-like interface, it has several special properties that make our design of the JAX version of CARFAC (CARFAC-JAX) deviate from the Matlab and NumPy versions, namely,
\begin{enumerate}
    \item JAX's transformations like autograd and JIT only operate on pure functions. Users are required to provide all the parameters needed for computation as function arguments. And modification on functions' input arguments needs to be avoided.
    \item Many JAX functions and transformations work on {\em pytrees} (basically containers of data). This inspired us to group different types of CARFAC parameters into different pytree classes.
\end{enumerate}

In the following, we explain the design in more details.

\subsection{Two Types of Functions and Four Types of Data Classes}

The functions implemented in CARFAC-JAX can be classified into two categories: design and initialization functions, and model functions. Design and initialization functions are used to design CARFAC's model parameters and initialize trainable weights. Their names have prefix like \texttt{design\_and\_init\_} or \texttt{design\_stage\_}. They are only invoked once at the beginning, before the emulation, and there is no need to JIT compile them. Consequently, there is no need to be concerned about execution speed or satisfying the requirements of \texttt{jax.jit} in modifying design functions. Conversely, model functions constitute the CARFAC's emulation process. They are usually compiled by \texttt{jax.jit}. Care should be taken to avoid slowing them down significantly. See Table \ref{tab:carfac-jax-example-functions} for example functions of each type.

\begin{table}
\centering
\begin{tabular}{ |c|c|p{7cm}| } 
\hline
{\bf Function type} & {\bf Examples} & {\bf Purposes} \\
\hline
\multirow{3}{2.1cm}{Design and Initialization} & \texttt{design\_and\_init\_filters()} & Design and initialize the CAR filters. \\ 
\cline{2-3}
& \texttt{design\_and\_init\_agc()} & Design and initialize the AGC step. \\
\cline{2-3}
& \texttt{design\_and\_init\_carfac()} & Design and initialize the whole CARFAC. \\ 
\hline
\multirow{3}{4em}{Model} & \texttt{car\_step()} & Implements the CAR step in emulation. \\
\cline{2-3}
& \texttt{ihc\_step()} & Implements the inner hair cell step in emulation. \\ 
\cline{2-3}
& \texttt{ohc\_nlf()} & Implements the outer hair cell non-linearity. \\ 
\hline
\end{tabular}
\caption{Examples of different types of functions in CARFAC-JAX.}
\label{tab:carfac-jax-example-functions}
\end{table}

We group the inputs and outputs of CARFAC-JAX into four types of pytree classes:   design parameters, hyper parameters, weights, and state. The motivation and purpose of each are:

\begin{enumerate}
    \item {\em Design parameters} are used for designing and initializing the CARFAC model. They are set manually by users and won't be changed in any CARFAC functions. They are only used in design and initialization and shouldn't be used in actual CARFAC model computation.
    \item {\em Hyper parameters} correspond to the hyper parameters in normal machine learning models. They are produced from the design processes based on the design parameters. Hyper parameters define the configurations of the model and their values shouldn't be changed in model computation or training. Therefore, they can be tagged as static in \texttt{jax.jit}, so we can use their values in Python conditionals and for-loops.
    \item {\em Weights} correspond to the trainable weights in neural network models. They are created and initialized by the design and initialization functions, and are used when the model functions are run. Their values shouldn't be changed in the model functions, but can be updated during training by optimizers.
    \item {\em State} contains the state variables of CARFAC. For the benefit of JAX, which mandates a purely functional programming style, they are passed as additional function inputs, and their new values are returned as function outputs. CARFAC's state updates at each sample time, representing the response of the filters, including those inside the CAR/AGC/OHC/IHC blocks, to sound input.
\end{enumerate}
Examples of different types of data structures can be found in Table \ref{tab:carfac-jax-data-structures}.

\begin{table}
\centering
\begin{tabular}{ |c|c|p{7cm}| } 
\hline
{\bf Data Structure} & {\bf Examples} & {\bf Meaning} \\
\hline
\multirow{3}{1.8cm}{Design Parameters} & \texttt{CarfacDesignParameters} & Used to design\&initialize the whole CARFAC. \\ 
\cline{2-3}
& \texttt{CarDesignParameters} & Used to design\&initialize the CAR step. \\
\cline{2-3}
& \texttt{AgcDesignParameters} & Used to design\&initialize the AGC step. \\ 
\hline
\multirow{3}{1.8cm}{Hyper parameters} & \texttt{CarfacHypers} & Hyperparameters of whole CARFAC. \\
\cline{2-3}
& \texttt{CarHypers} & Hyperparameters of the CAR step. \\ 
\cline{2-3}
& \texttt{AgcHypers} & Hyperparameters of the AGC step. \\ 
\hline
\multirow{3}{1.8cm}{Weights} & \texttt{CarfacWeights} & Trainable weights of whole CARFAC. \\
\cline{2-3}
& \texttt{CarWeights} & Trainable weights of the CAR step. \\ 
\cline{2-3}
& \texttt{AgcWeights} & Trainable weights of the AGC step. \\ 
\hline
\multirow{3}{1.8cm}{State} & \texttt{CarfacState} & State variables of whole CARFAC. \\
\cline{2-3}
& \texttt{CarState} & State variables of the CAR step. \\ 
\cline{2-3}
& \texttt{AgcState} & State variables of the AGC step. \\ 
\hline
\end{tabular}
\caption{Examples of different types of data structures in CARFAC-JAX.}
\label{tab:carfac-jax-data-structures}
\end{table}

% If we need to add usage example section, add it here.

\subsection{Benchmarks}
\label{sec:benchmark}

\begin{table}
    \centering
    \begin{tabular}{|c|c|c|c|c| }
    \hline
    {\bf Segment length} & {\bf MATLAB time} & {\bf NumPy time} & {\bf JAX time} & {\bf Speed ratio} \\
    \hline
    10 ms & N/A & 24.1 ms & 1.39 ms & 17.34 \\ \hline
    100 ms & N/A & 238 ms & 4.74 ms & 50.21\\ \hline
    1 s & N/A & 2.42 s  & 39.5 ms & 61.27 \\ \hline
    10 s & N/A & 24.1 s & 411 ms & 58.64 \\ \hline
    100 s & N/A & N/A & 4.43 s & N/A \\ \hline
    1000 s & N/A & N/A & 44.9 s & N/A \\ \hline
    1 s (split to 10 ms) & N/A & 2.40 s & 110 ms & 21.82 \\ \hline
    1 s (split to 100 ms) & N/A & 2.41 s & 46.2 ms & 52.16 \\ \hline
    2 s & 3.65 s & 4.80 s & 78.7 ms & 60.99 \\ \hline
    \end{tabular}
    \caption{Benchmark results of NumPy and JAX with speed factor comparison}
    \label{tab:jax_benchmark}
\end{table}

In the v2 update, both the NumPy and JAX versions include a simple benchmark suite that executes against segments of noise of various lengths. We've executed these benchmarks on several hardware configurations, and include results from one, which is an instance of an Intel {\it Xeon E5-2690 v4 @ 2.60GHz} CPU.

For the JAX version, we omit the time for JAX JIT compilation, which is roughly 800 milliseconds on systems we're running on. This is a once off cost for segments, and the compilation is cached in memory. All the benchmarks run against monaural 22050 Hz audio and the two-cap IHC model, which translate for other configurations. The resulting model is a 71 channel CARFAC model. We include the results in Table \ref{tab:jax_benchmark}.

We include a single benchmark taken from our Matlab tests to make a comparison and to validate that the JAX version is indeed much faster.

The JAX version approaches 61 times the speed of NumPy at longer segment lengths of audio, and the overall speed of JAX is a real-time factor (RTF) of approximately 0.04. For running on split segments of audio, simulating a streaming experience, the RTF is 0.11 with 10ms streaming chunks, or 0.046 with 100ms streaming chunks. We have not attempted to further improve the performance, nor include benchmarks running on many alternate execution platforms, such as GPUs or TPUs. 

With \texttt{use\_delay\_buffer} enabled, the JAX version runs up to 10\% faster on long segments (details not tabulated). 

The v1 C++/Eigen version, when running on similar workloads with a single CPU thread, is roughly twice as fast as the JAX version (details not tabulated).  And Eigen compiled with the \texttt{{openmp}} option and running 20 or more threads runs at least an order of magnitude faster still.  JAX, on the other hand, can compile to run on GPU and TPU systems, and supports automatic differentiation, so works better in big machine learning contexts.

On a MacBook Pro with Apple M2 Pro CPU running Matlab R2023B (which compiles to native code for Apple silicon), the Matlab CARFAC v2 runs almost four times faster than on the Intel/AMD CPUs we benchmarked with.  It takes a bit less than 1 second to process 2 seconds of monaural audio, with negligible additional overhead per segment. No significant difference was seen with  \texttt{use\_delay\_buffer} enabled.  The Activity Monitor shows about $10\%$ CPU utilization during execution, suggesting that it is running on a single core (of 12).

\section{Unchanged Misfeatures and Next Steps}

When we first created CARFAC, 22050 Hz seemed like a good compromise for a default sample rate, especially for speech processing.  But with its various nonlinearities creating distortion products that can alias, CARFAC does better with a higher over-sampling ratio, and some applications need more than speech bandwidth.  With computation being much less expensive now, sample rates of 48 kHz to 100 kHz might be preferable.  AMT experiments generally runs models at 100 kHz \cite{osses2022comparative}, and CARFAC does well with that.  Our current recommended sample rate is 48 kHz for modeling human hearing, but the code default remains 22050 Hz for compatibility.

The neural activity pattern (NAP) output of CARFAC is not as closely related to neural firing rates as we would like, especially in that it is zero in quiet and can go negative.  Subtracting off the mean in quiet (not shown in the diagrams in the book or here) is done so that the AGC filter average inputs and outputs will be 0 in quiet, making the damping in quiet equal to the \texttt{min\_damping} parameter.  We could move that subtraction out of the IHC and into the AGC to make the NAP nonnegative, but for compatibility have not done so.  To model neural firing rates more explicitly, we expect to develop a new version of the IHC with separate outputs for different spontaneous-rate classes of auditory nerve fibers, which will also be a place to incorporate parameterized synaptopathy, or ``hidden hearing loss'' \cite{monaghan2020hidden}, another mechanism of hearing impairment.  The effect of such changes on the CARFAC AGC's efferent feedback will need to be carefully considered.

It should be an easy job to port the v2 tests and changes to the C++/Eigen version, but we have not yet done so, as we are primarily building with JAX these days.  Volunteers are hereby solicited.

%\bibliographystyle{unsrt}

% %%% Comment out this section when you \bibliography{references} is enabled.

\end{document}